\documentclass{article}

\usepackage{arxiv}

\usepackage[utf8]{inputenc} 
\usepackage[T1]{fontenc}    
\usepackage{hyperref}       
\usepackage{url}            
\usepackage{booktabs}       
\usepackage{amsfonts}       
\usepackage{nicefrac}       
\usepackage{microtype}      
\usepackage{lipsum}
\usepackage{graphicx}
\usepackage{amsmath}
\graphicspath{ {./images/} }

\title{Richness Estimation with Species Identity Error}

\author{
  Jai-Hua Yen \\
  Department of Agronomy\\
  National Taiwan University\\
  Taipei 10617, Taiwan\\
  \texttt{r06621209@ntu.edu.tw} \\
   \And
 Chun-Huo Chiu \\
  Department of Agronomy\\
  National Taiwan University\\
  Taipei 10617, Taiwan\\
  \texttt{chchiu2017@ntu.edu.tw} \\
}

\begin{document}
\maketitle
\begin{abstract}
Richness estimation of an interesting area is always a challenge statistical work due to small sample size or species identity error. In the literatures, most richness estimators were only proposed to tackle the underestimation of the size-limited sample. However, species identity error almost occurs in each species survey and seriously reduces the accuracy of observed, singleton, and doubleton richness in turns to influence the behavior of richness estimator. Therefore, to estimate the true richness, the biased collected data due to species identity error should be modified before processing the richness estimation work. In the manuscript, we propose a new approach to correct the bias of richness estimation due to species identity error. First, a species list inventory from a subplot obtained by the investigator was used to estimate the species identity error rate. Then, we can correct the biased observed, singleton, and doubleton richness of the raw sampling data from the interesting area. Finally, the richness estimators proposed in the literatures could be supplied to get the more correct estimates based on adjusted observed data. To investigate the behavior of the proposed method, we performed simulations by generating data sets from various species models with different species identity error rates. For the purpose of illustration, the real data was supplied to demonstrate our proposed approach. A presence/absence of weeds species was surveyed in the organic farmland located at Soft Bridge County in the North of Taiwan.
\end{abstract}


\section{Introduction}
Long-term biodiversity monitoring is the basis for ecological research and promotion of organic agriculture. In recent years, more and more non-professional citizen scientists have participated in the projects of monitoring diversity, so the possibility of species identity errors may increase dramatically in the collected data. Therefore, correcting the impact of species identity error becomes an important statistical issue.

\quad Species richness is the most intuitive and widely used as biodiversity index due to its ecological intuitive concept and simplest form. However, due to the sampling limitation of time or other resources, completely species inventories in the wild field are almost unattainable goals. Therefore, the observed richness in the sample always underestimates the true species richness in the assemblage. In the literatures, among the discussed estimation approaches of species richness, the non-parametric methods are widely used in practical application, which include first order Jackknife approach, second order Jackknife approach by Burnham and Overton (1978) and Chao1 (or Chao2) lower bound estimator by Chao (1984) and Chao (1987). They all use the observed rare species in the sample (i.e. singletons and doubletons) to estimate the unseen richness in the sample. However, species identity error almost occurred in each survey especially in vegetation sampling, and it was ignored before and recently discussed in the literatures by Vittoz and Guisan (2007), Burg et al. (2015), and Morrison (2015). This identity error may seriously make observed richness biased and in turn the estimation of true richness will be seriously biased. Therefore, without error adjustment, the species richness estimation will be inaccurate based on original sampling data. In this manuscript, we have proposed a modify approach to revise the biased sampling data caused by species identity error. From the results of simulation study in session 3 show that our adjusting approach can revise the biased observed richness, singleton and doubleton richness. Also, the richness estimators based on the revised data effectively correct the bias caused by the species identity error.

\section{Methodology}
\label{sec:headings}
In this article, we choose Chao2 lower bound estimator for incidence data as our species richness estimator. Since we assume that species identity error exists in the process of sampling, adjustment of richness estimator should be considered.

\quad First, we need to estimate the mean species identity error rate of observer or investigator. Plant inventories from subplot of the area which the survey is conducted. We assume that the number of species ($S_{sub}$) and the categories of species in the subplot are known by the experiment designer but unknown by the observer who goes conducting inventories. After conducting inventories, we have the information that the number of observed species belongs to the subplot ($S_{sub,e}$) and the number of observed species does not exist in the subplot ($f_{sub,0}$).$\ X_i$ represents the record status of the survey of species $i$. When $X_i=1$, species $i$ has been recorded. When $X_i=0$, species $i$ has not been recorded. We assume the species identity error ($e$) is a random variable follows the distribution of $F(e)$ with mean $\bar{e}$. $r$ denotes the mean probability that a species is misidentified into another species which belongs to the sampling plot. $f_{sub,0}$ equals to the number of species which is misidentified and recorded as species do not exist in the subplot. Also, if plant inventories of the subplot are correct, then $S_{sub,e}$ should be equal to $S_{sub}$ species. However, when species identity error occurs, $S_{sub,e}$ may not be equal to $S_{sub}$ species. When the $i$-th species is misidentified and other species are not misidentified to the $i$-th species, $i$-th species is not recorded. After that, we have the equations:

\begin{equation}
    \begin{aligned}
        E\left(f_{sub,0}\right)&=\int{S_{sub}\times e\times\left(1-r\right)}dF\left(e\right)\\
        &\approx S_{sub}\times\bar{e}\times\left(1-r\right),
    \end{aligned}
\end{equation}

and

\begin{equation}
    \begin{aligned}
        E\left(S_{sub,e}\right)&=S_{sub}-\sum_{i=1}^{S_{sub}}E\left[I\left(X_i=0\right)\right]\\
        &\approx S_{sub}-S_{sub}\int{e\times\left(1-\frac{e}{\frac{S_{sub}}{r}-1}\right)^{S_{sub}-1}dF\left(e\right)}\\
        &\approx S_{sub}-S_{sub}\times\bar{e}\times\left(1-\frac{\bar{e}\times r}{S_{sub}-r}\right)^{S_{sub}-1}.
    \end{aligned}
\end{equation}

By solving those two equations, we have the estimate of $\bar{e}$ and $r$ which are denoted by $\widehat{\bar{e}}$ and $\hat{r}$.

\quad Second, the sampled observed, singleton, and doubleton richness should be adjusted after sampling in the plot. The true observed, singleton, and doubleton richness are denoted by $S_{obs}$, $Q_1$, and $Q_2$, respectively. The sampled observed, singleton, and doubleton richness without adjustment are denoted by $S_{obs,e}$, $Q_{1e}$, and $Q_{2e}$, respectively. When species identity error occurs, the sampled observed richness is formed by the observed species which do not misidentified and observed species which misidentified as species do not exist in the plot. Thus, we have the expected sampled observed richness:

\begin{equation*}
    E\left(S_{obs,e}\right)\approx E\left\{S_{obs}\left[\left(1-e\right)+e\times\left(1-r\right)\right]\right\}.
\end{equation*}

Next, we have the expected observed richness adjustment:

\begin{equation}
    S_{obs,a}=\frac{S_{obs,e}}{1-\widehat{\bar{e}}\times\hat{r}}.
\end{equation}

\quad When species identity error occurs, the possibilities of sampled singleton species are as follows: (1) singleton species which do not misidentified, and other species would not be misidentified as the singleton species at the same time, and (2) singleton species which misidentified as species do not exist in the plot, and other species would not be misidentified as the singleton species at the same time. Thus, we have the expected sampled singleton richness:

\begin{equation*}
    \begin{aligned}
    E\left(Q_{1e}\right)&\approx E\left\{Q_1\left[\left(1-e\right)+e\times\left(1-r\right)\right]\times\left(1-\frac{e}{\frac{S_{obs}}{r}-1}\right)^{S_{obs}-1}\right\} \\
    &\approx E\left\{Q_1\left[\left(1-e\right)+e\times\left(1-r\right)\right]\times e x p{\left(-e\times r\right)}\right\}.
    \end{aligned}
\end{equation*}

Similarly, when species identity error occurs, the possibilities of sampled doubleton species are as follows: (1) doubleton species which do not misidentified, and other species would not be misidentified as the singleton species at the same time, (2) doubleton species which misidentified as species do not exist in the plot, and other species would not be misidentified as the singleton species at the same time, and (3) when a singleton species misidentified to a singleton species, the doubleton richness increases by one unit, and other species would not be misidentified as the doubleton species which is formed by singleton species at the same time. Accordingly, we have the expected sampled doubleton richness:

\begin{equation*}
    \begin{aligned}
    E\left(Q_{2e}\right)&\approx E\left\{Q_2\left[\left(1-e\right)+e\times\left(1-r\right)\right]\times e x p{\left(-e\times r\right)}\right\}\\
    &+ E\left\{Q_1\times e\times r\times\left(1-\frac{1}{T}\right)\times\frac{Q_1}{S_{obs,a}}\times e x p{\left(-e\times r\right)}\right\}
    \end{aligned}
\end{equation*}

where $T$ denotes the number of sampling unit. By solving the two equations above, we have the singleton and doubleton richness adjustment:

\begin{equation}
    Q_{1a}=\frac{Q_{1e}}{(1-\widehat{\bar{e}}\times\hat{r})exp{\left(-\widehat{\bar{e}}\times\hat{r}\right)}},
\end{equation}

and

\begin{equation}
    Q_{2a}=\frac{Q_{2e}-Q_{1a}\times\widehat{\bar{e}}\times\hat{r}\times\left(1-\frac{1}{T}\right)\times\frac{Q_{1a}}{S_{obs,a}}\times e x p{\left(-\widehat{\bar{e}}\times\hat{r}\right)}}{(1-\widehat{\bar{e}}\times\hat{r})\times e x p{\left(-\widehat{\bar{e}}\times\hat{r}\right)}}.
\end{equation}

However, the estimation of traditional Chao2 estimator will be inaccurate even though $Q_{1a}$ and $Q_{2a}$ are asymptoticly unbiased. It causes the value of $\frac{Q_{1a}^2}{2Q_{2a}}$  overestimated. Hence, we choose first-order Jackknife and Chao2 richness estimator as the theoretical foundation of deriving the adjusted richness estimator. We propose an adjusted richness estimator by Taylor series expansion of  $E\left(\frac{Q_1^2}{2Q_2}\right)$  by the mean $Q_1$ and $Q_2$. Then we get the difference between  $\frac{[E(Q1)]^2}{2E(2Q2)}$  and  $E\left(\frac{Q_1^2}{2Q_2}\right)$  to have the adjust term:

\begin{equation*}
    E\left(\frac{Q_1^2}{2Q_2}\right)\approx\ \frac{\left[E\left(Q_1\right)\right]^2}{E\left(2Q_2\right)}+\frac{V\hat{a}r\left(Q_1\right)}{2E\left(Q_2\right)}-\frac{E\left(Q_1\right)C\hat{o}v\left(Q_1,Q_2\right)}{\left[E\left(Q_2\right)\right]^2}+\frac{\left[E\left(Q_1\right)\right]^2V\hat{a}r\left(Q_2\right)}{2\left[E\left(Q_2\right)\right]^3}
\end{equation*}

where $C\hat{o}v\left(Q_1,Q_2\right)=-\frac{Q_1Q_2}{\hat{S}}$, $V\hat{a}r\left(Q_i\right)=Q_i\left(1-\frac{Q_i}{\hat{S}}\right)$. Therefore, we have the adjusted richness estimator:

\begin{equation}
    {\hat{S}}_{adj}=S_{obs,a}+\frac{T-1}{T}max\left\{\left(\frac{Q_{1a}^2}{2Q_{2a}}-\frac{Q_{1a}}{2Q_{2a}}-\frac{Q_{1a}^2}{2Q_{2a}^2}\right),\ 0\right\}
\end{equation}

When $0\le Q_{2a}\le1$, by simulation studies, the adjusted richness estimator will be:

\begin{equation}
    {\hat{S}}_{adj}=S_{obs,a}+\frac{T-1}{T}Q_{1a}
\end{equation}

\section{Result}
\subsection{Simulation Results}
To test the performance of the adjusted richness estimator, we presented the simulation results by several species detection models and different settings of number of sampling units. We fixed $S_{sub}=40$ and $S=100$. 500 simulation data sets were generated and 200 bootstrapping trials were conducted by each simulation data. The bootstrapping method is regenerating $S_{obs,a}$, $Q_{1a}$, and $Q_{2a}$ by binomial distribution independently in order to increase the estimated standard error while the traditional bootstrapping method usually underestimates the standard error in this case. In true method, the estimation of species richness used the traditional Chao2 estimator by the data without species identity error. In observed method, the estimation of species richness used the traditional Chao2 estimator by the data with species identity error. In adjusted method, the estimation of species richness used the adjusted richness estimator by the data with species identity error.

\quad When species identity error occurs, the estimate of species richness by observed method will be underestimated, which causes larger bias. The large bias still exists even though the increase of the number of sampling units. Since adjusted method slightly overestimated species richness when the species identity error rate is large, it reduces a great quantity of bias. The variation of observed method is lower, and it remains the same by different species identity error rate. The adjusted method has a higher variation. When species identity error rate is larger, the variation of adjusted method is larger. By evaluating both bias and variation, the observed method has a larger RMSE (Root Mean Square Error) due to its larger bias. The adjusted method has about half RMSE of the observed method when the number of sampling unit is large.


\begin{table}[ht]
\centering
\caption{Comparison of species richness estimator for incidence data based on 500 simulation data sets and 200 bootstrapping trials under random uniform (0, 1) model, with $\bar{p}=0.51$, $CV=0.53$, $S=100$, $S_{sub}=40$, $T=5$, and $r=0.91$.}
\begin{tabular}[t]{lccccccccccc}
\hline
$\bar{e}$ & $\widehat{\bar{e}}$ & Method & $S_{obs}$ & $Q_1$ & $Q_2$ & $\hat{S}$ & Bias & s.e. & $\hat{s.e.}$ & RMSE &\\
\hline
0     & 0     & True     & 85.2 & 15.3 & 17.3 & 91.37 & -8.63  & 4.82  & 4.19  & 9.89   \\
0.053 & 0.058 & Observed & 81.5 & 13.9 & 15.8 & 87.22 & -12.78 & 5.46  & 4.06  & 13.9   \\
      &       & Adjusted & 86.3 & 15.6 & 17.5 & 92.05 & $-7.95^\star$ & 7.17  & 8.33  & $10.71^\dagger$ \\
0.097 & 0.098 & Observed & 78.3 & 13.2 & 14.8 & 83.72 & -16.28 & 5.29  & 3.95  & 17.12  \\
      &       & Adjusted & 86.3 & 15.9 & 17.5 & 92.2  & $-7.8^\star$  & 7.92  & 9.4   & $11.12^\dagger$ \\
0.15  & 0.157 & Observed & 74   & 11.7 & 13.4 & 78.86 & -21.14 & 5.24  & 3.75  & 21.78  \\
      &       & Adjusted & 86.8 & 16   & 17.6 & 92.89 & $-7.11^\star$ & 10.33 & 10.2  & $12.54^\dagger$ \\
0.199 & 0.209 & Observed & 70.7 & 10.3 & 12.7 & 74.71 & -25.29 & 5.01  & 3.34  & 25.78  \\
      &       & Adjusted & 88.3 & 15.8 & 18.5 & 94.34 & $-5.66^\star$ & 14.05 & 11.12 & $15.15^\dagger$\\
\hline
\end{tabular}
\\
$^\star$ Denotes the smaller bias. $^\dagger$ Denotes the smaller RMSE.
\end{table}%


\begin{table}[ht]
\centering
\caption{Comparison of species richness estimator for incidence data based on 500 simulation data sets and 200 bootstrapping trials under random uniform (0, 1) model, with $\bar{p}=0.51$, $CV=0.53$, $S=100$, $S_{sub}=40$, $T=20$, and $r=0.91$.}
\begin{tabular}[t]{lccccccccccc}
\hline
$\bar{e}$ & $\widehat{\bar{e}}$ & Method & $S_{obs}$ & $Q_1$ & $Q_2$ & $\hat{S}$ & Bias & s.e. & $\hat{s.e.}$ & RMSE &\\
\hline
      &       & True     & 95.3 & 4.1 & 3.9 & 98.8  & -1.2   & 4.9   & 4.25 & 5.06   \\
0.053 & 0.055 & Observed & 91.2 & 3.9 & 3.6 & 94.8  & -5.2   & 5.46  & 4.45 & 7.53   \\
      &       & Adjusted & 96.1 & 4.3 & 4   & 97.85 & $-2.15^\star$ & 5.26  & 5.39 & $5.68^\dagger$  \\
0.097 & 0.095 & Observed & 87.3 & 3.3 & 3.5 & 90.1  & -9.9   & 5.15  & 3.76 & 11.15  \\
      &       & Adjusted & 95.8 & 4   & 4.1 & 97.1  & $-2.9^\star$  & 6.52  & 5.72 & $7.14^\dagger$  \\
0.15  & 0.151 & Observed & 82.9 & 3.1 & 2.9 & 85.61 & -14.39 & 5.21  & 3.79 & 15.31  \\
      &       & Adjusted & 96.7 & 4.1 & 3.9 & 97.94 & $-2.06^\star$ & 8.94  & 6.23 & $9.17^\dagger$  \\
0.199 & 0.21  & Observed & 79.2 & 2.9 & 2.7 & 81.79 & -18.21 & 5.25  & 3.66 & 18.95  \\
      &       & Adjusted & 98.8 & 4.4 & 4   & 100.5 & $0.46^\star$  & 11.52 & 7.04 & $11.53^\dagger$ \\
\hline
\end{tabular}
\\
$^\star$ Denotes the smaller bias. $^\dagger$ Denotes the smaller RMSE.
\end{table}%

\newpage


\begin{table}[ht]
\centering
\caption{Comparison of species richness estimator for incidence data based on 500 simulation data sets and 200 bootstrapping trials under $(0.8\times Uniform\left(0.1,\ 0.3\right)+0.2\times Uniform\left(0.4,\ 1\right))$, with $\bar{p}=0.29$, $CV=0.7$, $S=100$, $S_{sub}=40$, $T=5$, and $r=0.67$.}
\begin{tabular}[t]{lccccccccccc}
\hline
$\bar{e}$ & $\widehat{\bar{e}}$ & Method & $S_{obs}$ & $Q_1$ & $Q_2$ & $\hat{S}$ & Bias & s.e. & $\hat{s.e.}$ & RMSE &\\
\hline
     &       & True     & 72   & 32.4 & 19.8 & 94.98 & -5.02  & 11.38 & 10.76 & 12.44  \\
0.053 & 0.056 & Observed & 69.7 & 30.4 & 19   & 90.91 & -9.09  & 11.13 & 10.22 & 14.37  \\
      &       & Adjusted & 72.5 & 32.9 & 19.9 & 94.7  & $-5.3^\star$  & 12.56 & 12.98 & $13.63^\dagger$ \\
0.097 & 0.1   & Observed & 67.3 & 28.8 & 18.3 & 87.32 & -12.68 & 11.12 & 9.91  & $16.87^\dagger$ \\
      &       & Adjusted & 72.3 & 33.1 & 19.8 & 95.78 & $-4.22^\star$ & 21.14 & 15.12 & 21.56  \\
0.15  & 0.155 & Observed & 64.7 & 26.4 & 17.7 & 82.27 & -17.73 & 11.77 & 9.06  & $21.28^\dagger$ \\
      &       & Adjusted & 72.7 & 33.1 & 20.1 & 96.26 & $-3.74^\star$ & 21.81 & 17.28 & 22.13  \\
0.199 & 0.203 & Observed & 63.1 & 24.9 & 17.2 & 78.81 & -21.19 & 9.08  & 8.36  & 23.06  \\
      &       & Adjusted & 73.9 & 33.9 & 20.3 & 98.02 & $-1.98^\star$ & 22.62 & 19.58 & $22.71^\dagger$\\
\hline
\end{tabular}
\\
$^\star$ Denotes the smaller bias. $^\dagger$ Denotes the smaller RMSE.
\end{table}%


\begin{table}[ht]
\centering
\caption{Comparison of species richness estimator for incidence data based on 500 simulation data sets and 200 bootstrapping trials under $(0.8\times Uniform\left(0.1,\ 0.3\right)+0.2\times Uniform\left(0.4,\ 1\right))$, with $\bar{p}=0.29$, $CV=0.7$, $S=100$, $S_{sub}=40$, $T=20$, and $r=0.67$.}
\begin{tabular}[t]{lccccccccccc}
\hline
$\bar{e}$ & $\widehat{\bar{e}}$ & Method & $S_{obs}$ & $Q_1$ & $Q_2$ & $\hat{S}$ & Bias & s.e. & $\hat{s.e.}$ & RMSE &\\
\hline
      &       & True     & 97.8 & 7   & 11.9 & 100.25 & 0.25   & 2.56 & 2.43 & 2.57  \\
0.053 & 0.056 & Observed & 94.7 & 6.6 & 11.1 & 97.08  & -2.92  & 2.98 & 2.37 & $4.17^\dagger$ \\
      &       & Adjusted & 98.5 & 7.1 & 12   & 100.62 & $0.62^\star$  & 4.55 & 5.8  & 4.59  \\
0.097 & 0.102 & Observed & 91.5 & 6.2 & 10.4 & 93.78  & -6.22  & 3.72 & 2.34 & 7.25  \\
      &       & Adjusted & 98.6 & 7.2 & 12   & 100.76 & $0.76^\star$  & 6.24 & 6.97 & $6.29^\dagger$ \\
0.15  & 0.151 & Observed & 88.2 & 5.8 & 9.8  & 90.42  & -9.58  & 3.69 & 2.31 & 10.27 \\
      &       & Adjusted & 98.5 & 7.2 & 12.1 & 100.62 & $0.62^\star$  & 7.5  & 7.5  & $7.53^\dagger$ \\
0.199 & 0.204 & Observed & 85.4 & 5.4 & 9.1  & 87.45  & -12.55 & 4.2  & 2.3  & 13.24 \\
      &       & Adjusted & 99.9 & 7.3 & 12.2 & 102.08 & $2.08^\star$  & 9.64 & 7.98 & $9.86^\dagger$ \\
\hline
\end{tabular}
\\
$^\star$ Denotes the smaller bias. $^\dagger$ Denotes the smaller RMSE.
\end{table}%

\subsection{Real Data Analysis}
The data set was collected of weed species from organic farmland located at Soft Bridge county in the North of Taiwan. There are 12 transect lines with length 20m each were conducted. Only the incidence (detection or non-detection) of species in each transect line was recorded. Before richness estimation, a subplot occupied by 40 known weed species was treated as the testing of the degree of investigator's skill. Compare these 40 weed species list with the inventories of the investigator, we have $S_{sub}=40$, $S_{sub,e}=35$, and $f_{sub,0}=1$. Therefore, we have the estimate of $\widehat{\bar{e}}=0.14$ and $\hat{r}=0.82$ based on equations (1) and (2). Many of the misidentified species were misidentified as species which did not exist in the plot. The summary of the frequency counts of weed species is in Table 5. The result using our adjusted estimator is in Table 6. By simulation studies, the error rate is high in this case. Hence, the estimate of species richness using row data directly underestimates and the adjusted estimator should be applied to get the accurate estimate of species richness.


\begin{table}[ht]
\centering
\caption{Summary of the data set of weed species frequency counts at Soft Bridge county in the North of Taiwan, with $T=12$.}
\begin{tabular}[t]{lccccccccccccc}
\hline
Frequency & $Q_1$ & $Q_2$ & $Q_3$ & $Q_4$ & $Q_5$ & $Q_6$ & $Q_7$ & $Q_8$ & $Q_9$ & $Q_{10}$ & $Q_{11}$ & $Q_{12}$\\
\hline
Counts & 18 & 9 & 12 & 8 & 6 & 4 & 1 & 4 & 3 & 3 & 2 & 3 \\
\hline
\end{tabular}
\end{table}%


\begin{table}[ht]
\centering
\caption{Species richness adjustment for data set of weed species from Soft Bridge county in the North of Taiwan in farmland, with $T=12$, $\hat{r}=0.82$, and $\widehat{\bar{e}}=0.14$.}
\begin{tabular}[t]{lccccccccccccc}
\hline
Method & $S_{obs}$ & $Q_1$ & $Q_2$ & $\hat{S}$ & $\hat{s.e.}$\\
\hline
Observed & 74.0 & 19.0 & 9.0  & 92.4  & 11.27 \\
Adjusted & 83.6 & 24.1 & 10.6 & 105.4 & 18.68 \\
\hline
\end{tabular}
\end{table}%

\section{Discussion and Conclusion}
Species richness is the simplest and most popular measure of biodiversity. The approach of estimating species richness is widely discussed due to its application in many ecological or agricultural issues mentioned by Carvalheiro et al. (2011) and Garibaldi et al. (2013). In the manuscript, we demonstrated the effect of species identity error while sampling in estimating species richness. When the mean probability that a species is misidentified into another species which belongs to the sampling plot is high, the observed richness and singleton richness will be seriously negative biased which implying most richness estimators’ serious underestimation even though increasing sampling units. Our simulations show that the adjusted richness estimator removes a large proportion of the negative bias under different settings of sampling units, species identity error, and species detection model. We suggest that the adjusted richness estimator for incidence data should be applied to estimate species richness of the target region since species identity error occurs almost in every investigation of species.

\section{Acknowledgements}
The research was supported by the Taiwan National Science Council under Project 107-2118-M-002-001-MY2 and Council of Agriculture under Project 107AS-1.2.7-ST-a6.


\end{document}